\documentclass[useAMS,usenatbib]{mn2e}

\usepackage{graphics}
\usepackage{epsf}
\usepackage{subfigure}

\def\ni{\noindent}
\def\apj{Astrophysics Journal}
\def\aj{Astronomical Journal}
\def\mnras{Monthly Notices of the Royal Astronomical Society}
\def\aap{Astronomy and Astrophysics}

\title[Cepheid P-L relation from {\it AKARI}]{Cepheid Period-Luminosity Relation from the {\it AKARI} Observations}

\author[Ngeow et al.]{Chow-Choong Ngeow$^{1}$\thanks{E-mail: cngeow@astro.ncu.edu.tw}, Yoshifusa Ita$^{2}$, Shashi M. Kanbur$^{3}$, Hilding Neilson$^{4}$,
\and Takashi Onaka$^{5}$ and Daisuke Kato$^{5}$ 
\\
$^{1}$Graduate Institute of Astronomy, National Central University, Jhongli City, 32001, Taiwan (R.O.C.)\\
$^{2}$National Astronomical Observatory of Japan, 2-21-1 Osawa, Mitaka, Tokyo, 181-8588, Japan \\
$^{3}$Department of Physics, SUNY Oswego, Oswego, NY 13126, USA \\
$^{4}$Department of Astronomy \& Astrophysics, University of Toronto, Toronto, ON, M5S 3H8, Canada \\
$^{5}$Department of Astronomy, Graduate School of Science, The University of Tokyo, Bunkyo-ku, Tokyo 113-0033, Japan
}

\begin{document}

\date{Accepted 2010 month date. Received 2009 September 09; in original form 2009 September 01}

\pagerange{\pageref{firstpage}--\pageref{lastpage}} \pubyear{2009}

\maketitle

\label{firstpage}

\begin{abstract}

In this paper, we derive the period-luminosity (P-L) relation for Large Magellanic Cloud (LMC) Cepheids based on mid-infrared {\it AKARI} observations. {\it AKARI's IRC} sources were matched to the OGLE-III LMC Cepheid catalog. Together with the available $I$ band light curves from the OGLE-III catalog, potential false matches were removed from the sample. This procedure excluded most of the sources in the $S7$ and $S11$ bands: hence only the P-L relation in the $N3$ band was derived in this paper. Random-phase corrections were included in deriving the P-L relation for the single epoch {\it AKARI} data, even though the derived P-L relation is consistent with the P-L relation without random-phase correction, though there is a $\sim7$ per-cent improvement in the dispersion of the P-L relation. The final adopted $N3$ band P-L relation is $N3 = -3.246\log(P)+15.844$, with a dispersion of $0.149$.

\end{abstract}


\section{Introduction}

The mid-infrared Cepheid period-luminosity (P-L, also known as the Leavitt Law) relation is becoming more important in the future extra-galactic distance scale studies \citep[for example, see][]{fre08a}. This is mainly because the effect of extinction is negligible in the mid-infrared \citep{fre08b,nge09}. Other advantages of using the mid-infrared P-L relation include \citep[see also the discussion in][]{fre08b,nge08,mad09,nge09,mar09}: (a) the dispersion of the infrared P-L relation is smaller compared to the $BVI$ P-L relations; (b) the mid-infrared light curves are expected to have smaller amplitudes than the optical counterparts, therefore a smaller number (or even a single epoch) of observations are adequate to derive accurate mean magnitudes and P-L relations, and (c), metallicity may not affect the luminosity of Cepheids at these wavelengths \citep[however, metallicity may affect the {\it observed} magnitudes due to mass loss, see][]{nei09}. 

The SAGE project \citep[Surveying the Agents of a Galaxy's Evolution,][]{mei06} has surveyed the Large Magellanic Cloud (LMC) using the {\it Spitzer's IRAC} instrument at two epochs. The first data released from the SAGE team only included Epoch 1 data. This has been used by \citet{nge08} and \citet{fre08b} to derive the LMC P-L relation in {\it IRAC} bands. The second release of the SAGE data contains both the Epoch 1 and Epoch 2 data, and were used to improve the {\it IRAC} band P-L relations in \citet{nge09} and \citet{mad09}, respectively. The main difference between the Freedman/Madore group and our group is that they matched the SAGE catalogs to the LMC Cepheids from \citet{per04}, while we adopted the Cepheid catalogs released from the OGLE team \citep[Optical Gravitational Lensing Experiment,][]{uda99,sos08}. The two approaches lead to a discrepancy found in the slopes of the {\it IRAC} band P-L relations, in a sense that the slopes found by Freedman/Madore group are steeper than the slopes found by our group (see Table \ref{tab_compare} for a comparison of the slopes found by these two groups). Nevertheless, the SAGE catalogs used by both groups do not contain information regarding the time of observation. Hence, random-phase corrections cannot be applied to the single or double epoch SAGE data, in order to derive the mean magnitudes in {\it IRAC} bands. 

\begin{table}
  \centering
  \caption{Comparison of the P-L slopes in {\it Spitzer's IRAC} bands.}
  \label{tab_compare}
  \begin{tabular}{lcc} \hline
    Band & \citet{nge09} & \citet{mad09} \\
    \hline 
    $3.6\mu\mbox{m}$ & $-3.253\pm0.010$ & $-3.40\pm0.02$ \\
    $4.5\mu\mbox{m}$ & $-3.214\pm0.010$ & $-3.35\pm0.02$ \\
    $5.8\mu\mbox{m}$ & $-3.182\pm0.020$ & $-3.44\pm0.03$ \\
    $8.0\mu\mbox{m}$ & $-3.197\pm0.036$ & $-3.49\pm0.03$ \\    
    \hline
  \end{tabular}
\end{table}

Independent of the SAGE project, the {\it AKARI} satellite \citep{mur07} has surveyed the LMC using the on board {\it IRC} \citep[Infrared Camera,][]{ona07} instrument. Initial results were published in \citet{ita08}. The wavelength coverage of {\it AKARI's IRC} instrument is similar to {\it Spitzer's IRAC} bands. Therefore, the main purpose of this paper is to construct P-L relations based on {\it AKARI} observations, and attempt to resolve the discrepancy in the slopes of the {\it IRAC} band P-L relations calculated by our group and the Madore/Freedman group. The {\it AKARI} internal catalog includes the time of observation: this allows us to obtain the phase information of the matched Cepheids:

\begin{eqnarray}
  \phi & = & \mathrm{mod}(\frac{t-t_0}{P}), 
\end{eqnarray}

\ni where $t$ is the time of observation, $t_0$ is the time of maximum light (either in $V$ or $I$ band) and $P$ is the pulsational period. Hence $\phi\in [0,1]$ represents one pulsation cycle of the Cepheid. The phase information allow us to apply random-phase corrections and derive mean magnitudes: such an approach has not been applied to the SAGE data. In Sections 2 and 3 we describe the data used in this study and our adopted methodology for random-phase correction. We present the results and discussion in Section 4, and followed by the conclusion in Section 5. Extinction correction is ignored in this paper since it is negligible in the mid-infrared bands.

\section{The Data}

The {\it AKARI} catalog provides photometric data at $3$ ($N3$), $7$ ($S7$), $11$ ($S11$), $15$ ($L15$), and $24$ ($L24$) microns together with their time of observation. Photometry is on the {\it IRC}-Vega magnitude system, which is defined in \citet{tan08}. Details of the data reduction and catalog compilation processes are described in \citet{ita08}, and will not be repeated here. The coordinates given in the {\it AKARI} catalog are calculated by matching detected point sources with corresponding 2MASS sources. If matching with the 2MASS catalog is unsuccessful (such cases usually occur in $L15$ and $L24$ images), then we use the SAGE point source catalog \citep{mei06} as the positional reference. The root-mean-squares of the residuals between the input 2MASS/SAGE catalog coordinates and the fitted coordinates are smaller than $1.2$ arc-second for $N3$, $2.6$ arc-second for $S7$ and $S11$, and $2.9$ arc-second for $L15$ and $L24$, respectively. The {\it AKARI} catalog coordinates should be accurate to that extent relative to the 2MASS and SAGE catalog coordinates.

\begin{figure}
  \centering 
  \epsfxsize=8.0cm{\epsfbox{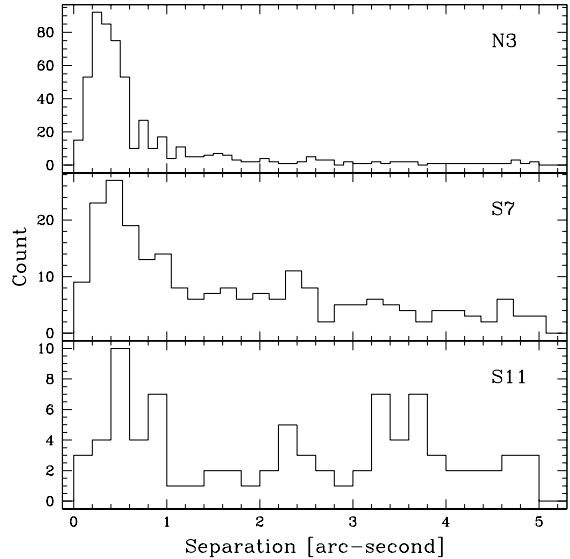}}
  \caption{Distributions of the separation between the matched {\it AKARI} sources and the input OGLE-III LMC Cepheids.}
  \label{separation}
\end{figure}

\begin{figure}
  \centering 
  \epsfxsize=8.0cm{\epsfbox{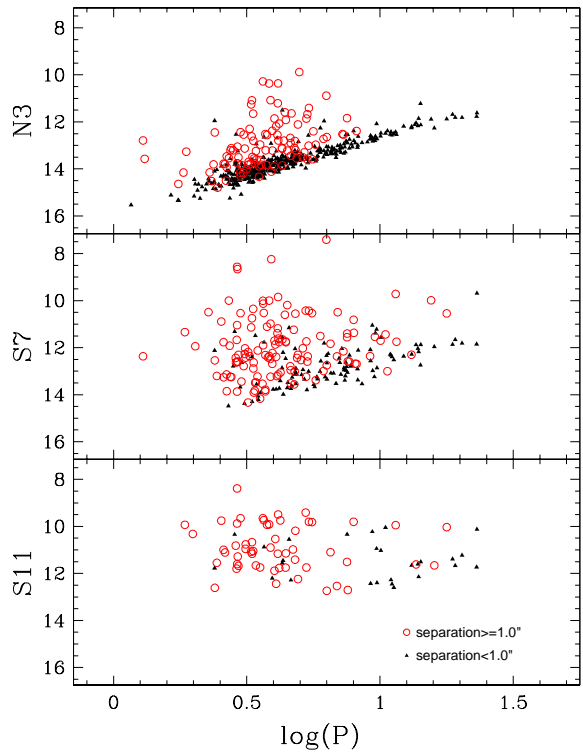}}
  \caption{The P-L relations constructed from all of the matched {\it AKARI} sources to the input OGLE-III LMC Cepheids, with a crude division of the matched sources that has separation greater (open circles) and smaller (filled triangles) than $1$ arc-second. Error bars are omitted for clarity.}
  \label{pl_raw}
\end{figure}

We matched the {\it AKARI's IRC} point source catalog given in \citet{ita08}\footnote{A revision of this catalog is currently in progress (Kato et al., 2010, in preparation), and will be published in the near future. Preliminary analysis shows that the difference of the $N3$ band photometry is negligible between the current and revised catalogs.} to the OGLE-III fundamental mode Cepheid catalog from \citet{sos08}. Due to the smaller area coverage of the {\it AKARI} survey \citep[see Figure 1 of][]{ita08} as compared to the SAGE project, the number of matched sources is reduced to $537$, $226$ and $83$ in the {\it AKARI's} $N3$, $S7$ and $S11$ band, respectively. Figure \ref{separation} shows the distribution of the separation between matched {\it AKARI} sources and the OGLE-III LMC Cepheids. The numbers of matched sources with separation greater than $1$ arc-second are $100$ (18.6\%), $124$ (54.9\%) and $55$ (63.3\%) in the $N3$, $S7$ and $S11$ band, respectively. The corresponding P-L relations from all the matched sources are presented in Figure \ref{pl_raw}. This Figure shows that there is a well defined sequence of P-L relation in $N3$ band. The P-L relation sequence is not obvious in the $S7$ band and almost disappears in the $S11$ band. A fraction of the matched sources deviate from the expected P-L relation sequence, with most of them having a separation greater than $1$ arc-second. They are clearly the false matched {\it AKARI} sources to the OGLE-III Cepheid catalog. 

\begin{figure*}
  \vspace{0cm}
  \hbox{\hspace{-0.2cm}
    \epsfxsize=6.0cm \epsfbox{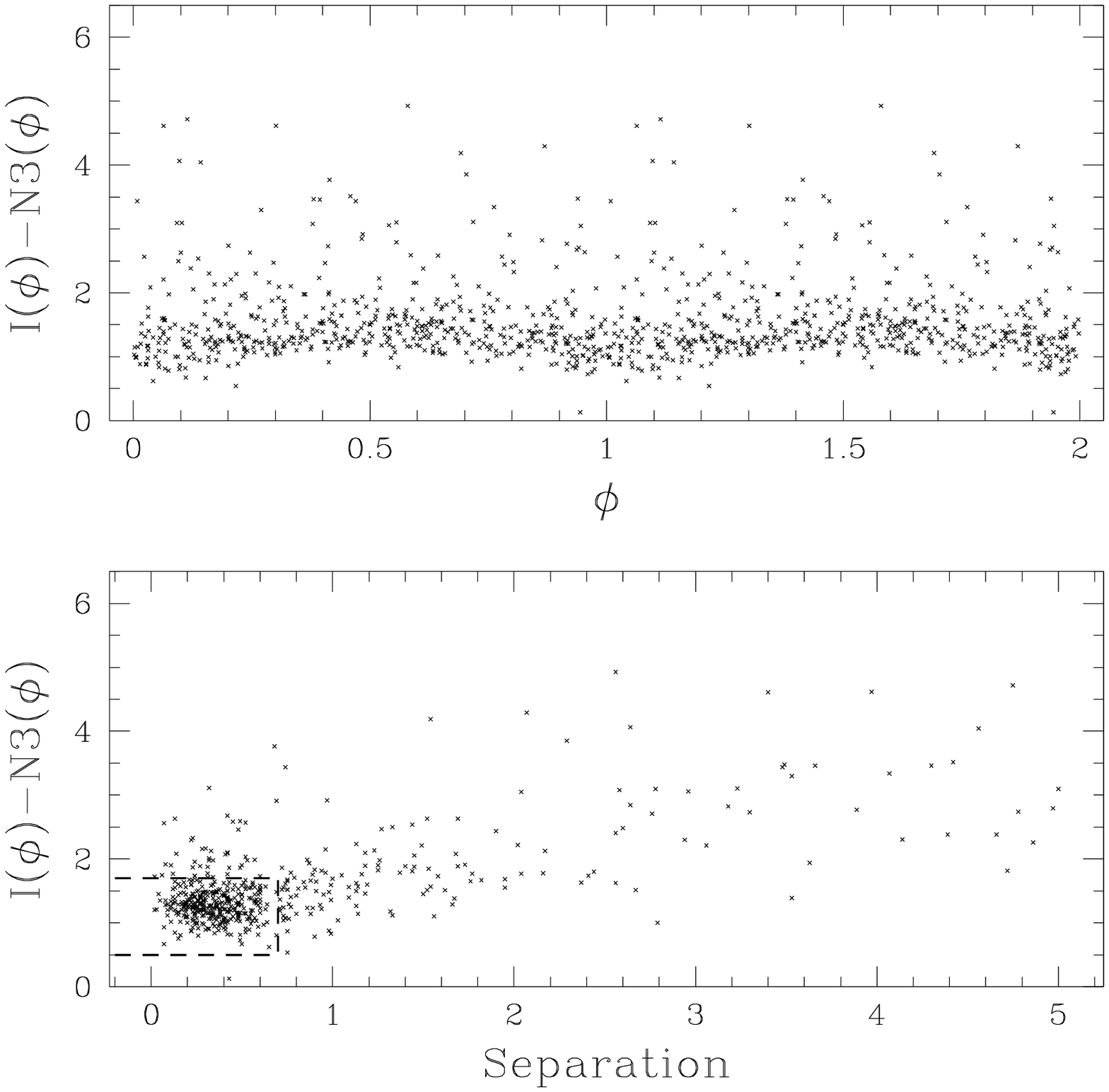}
    \epsfxsize=6.0cm \epsfbox{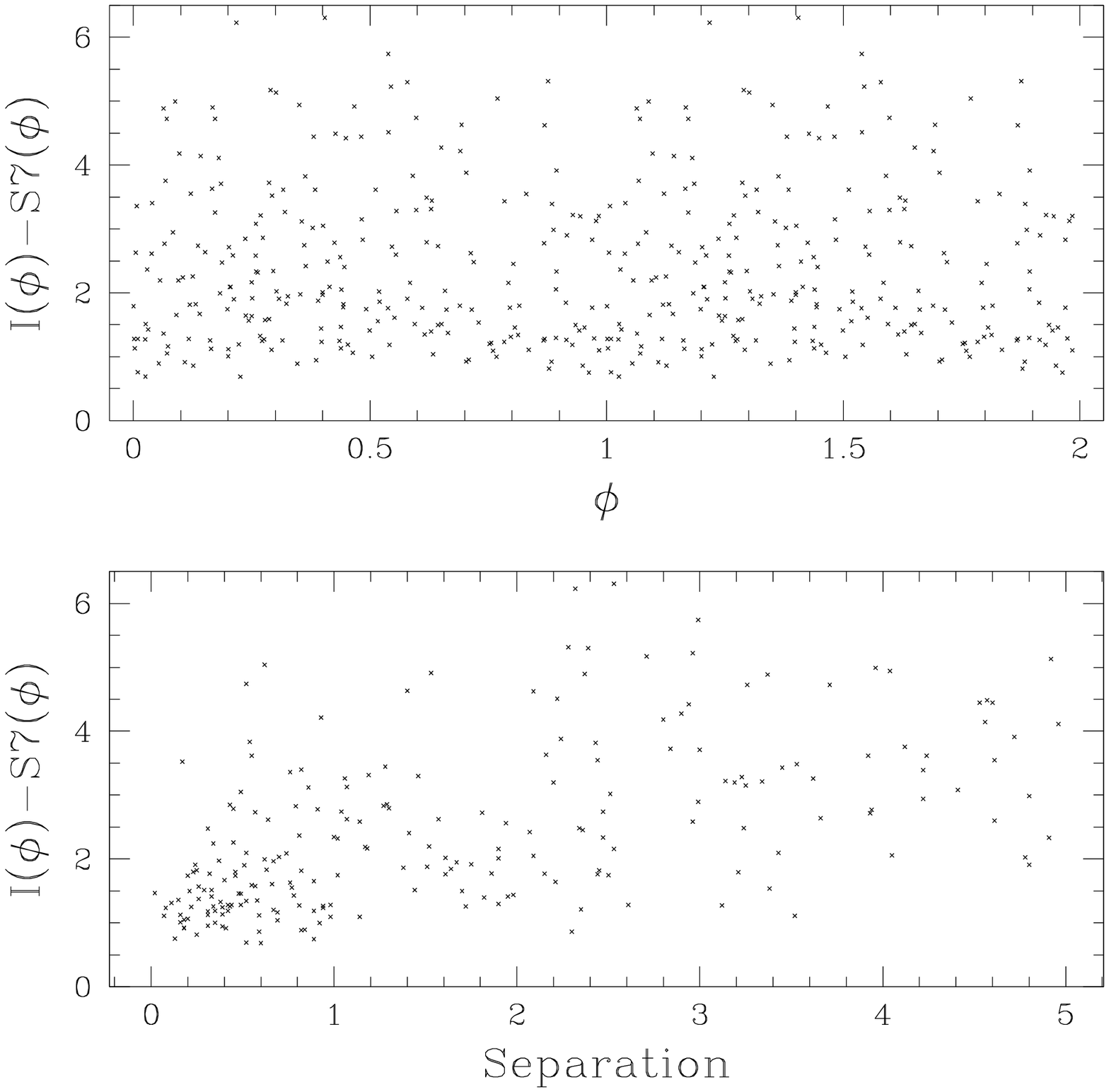}
    \epsfxsize=6.0cm \epsfbox{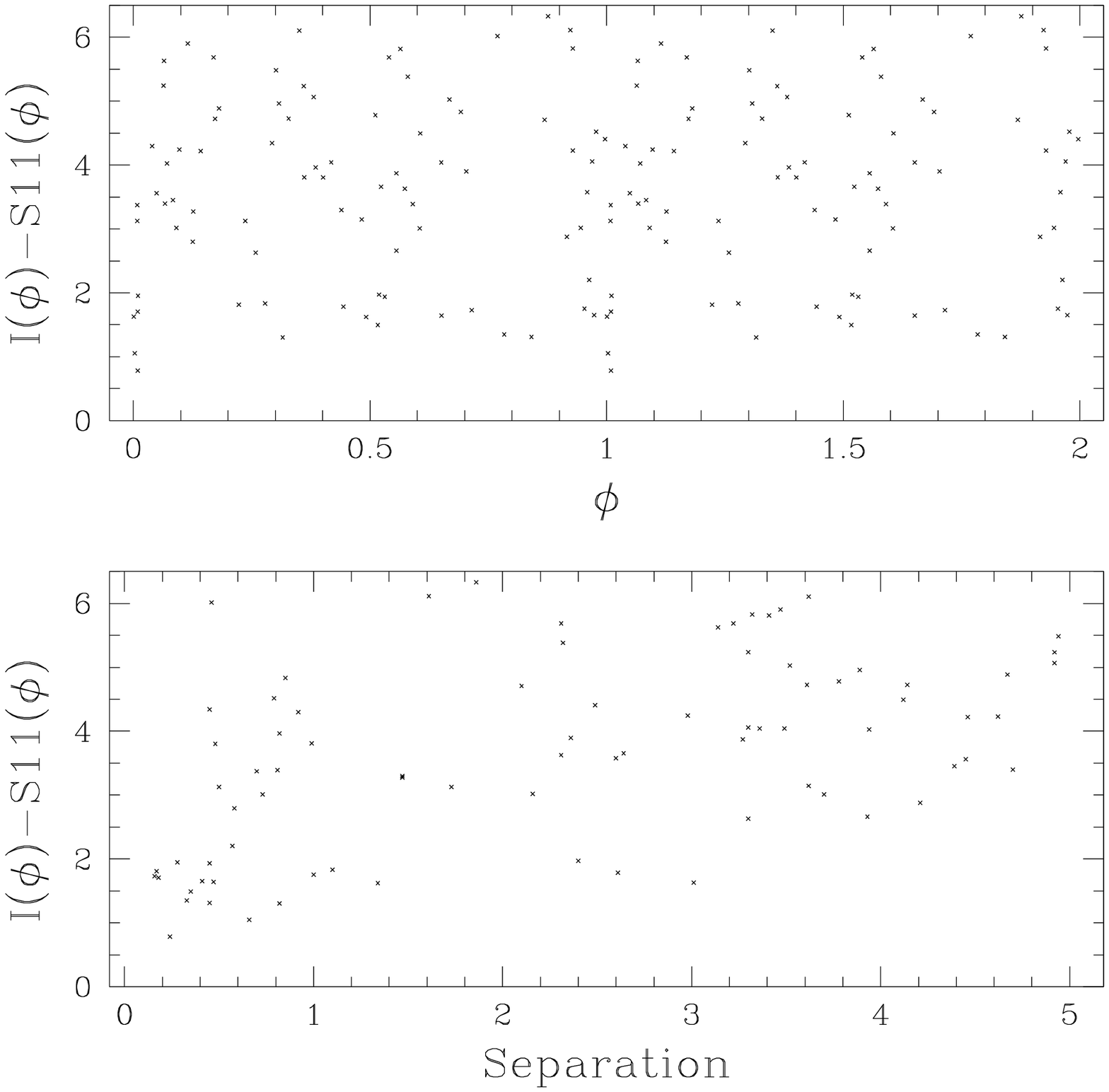}
  }
  \vspace{0cm}
  \caption{The $I(\phi)-AKARI(\phi)$ colours of the matched {\it AKARI} sources to the OGLE-III Cepheids as a function of phases (top panels) and separations (bottom panels), for the $N3$ (left pane), $S7$ (middle panel) and $S11$ (right panel). The dashed lines in lower left panel are the adopted cuts to the data (see text for details). Error bars are omitted for clarity.}
  \label{rawcolor}
\end{figure*}

\subsection{Selection of Data}

Since the phase information for the matched {\it AKARI} sources can be obtained from equation (1), we can construct the colour of the matched sources using the full $I$ band light curves from OGLE-III catalog given in \citet{sos08}. We match the phased $I$ band magnitude that is closest to the phase of the {\it AKARI} magnitudes in all three bands, and construct the $I(\phi)-AKARI(\phi)$ colour at a given phase $\phi$ for a particular matched source. The phase difference between the $I$ band magnitudes and the {\it AKARI} sources is always less than $0.03$, with majority of them smaller than $0.003$. We then plotted the $I(\phi)-AKARI(\phi)$ colour as a function of phases and separations in Figure \ref{rawcolor}.

\begin{figure*}
  \vspace{0cm}
  \hbox{\hspace{0.5cm}
    \epsfxsize=8.0cm \epsfbox{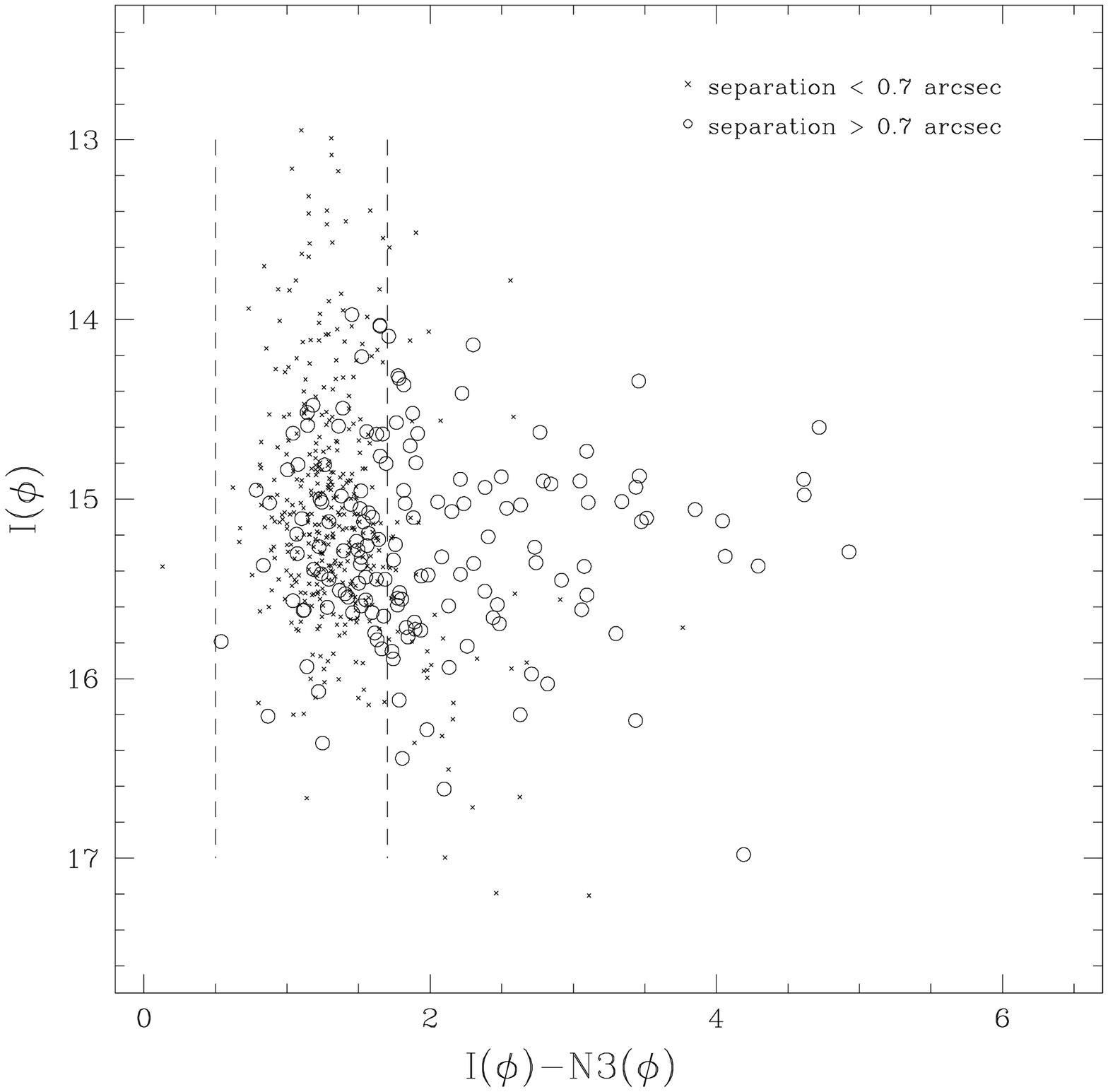}
    \epsfxsize=8.0cm \epsfbox{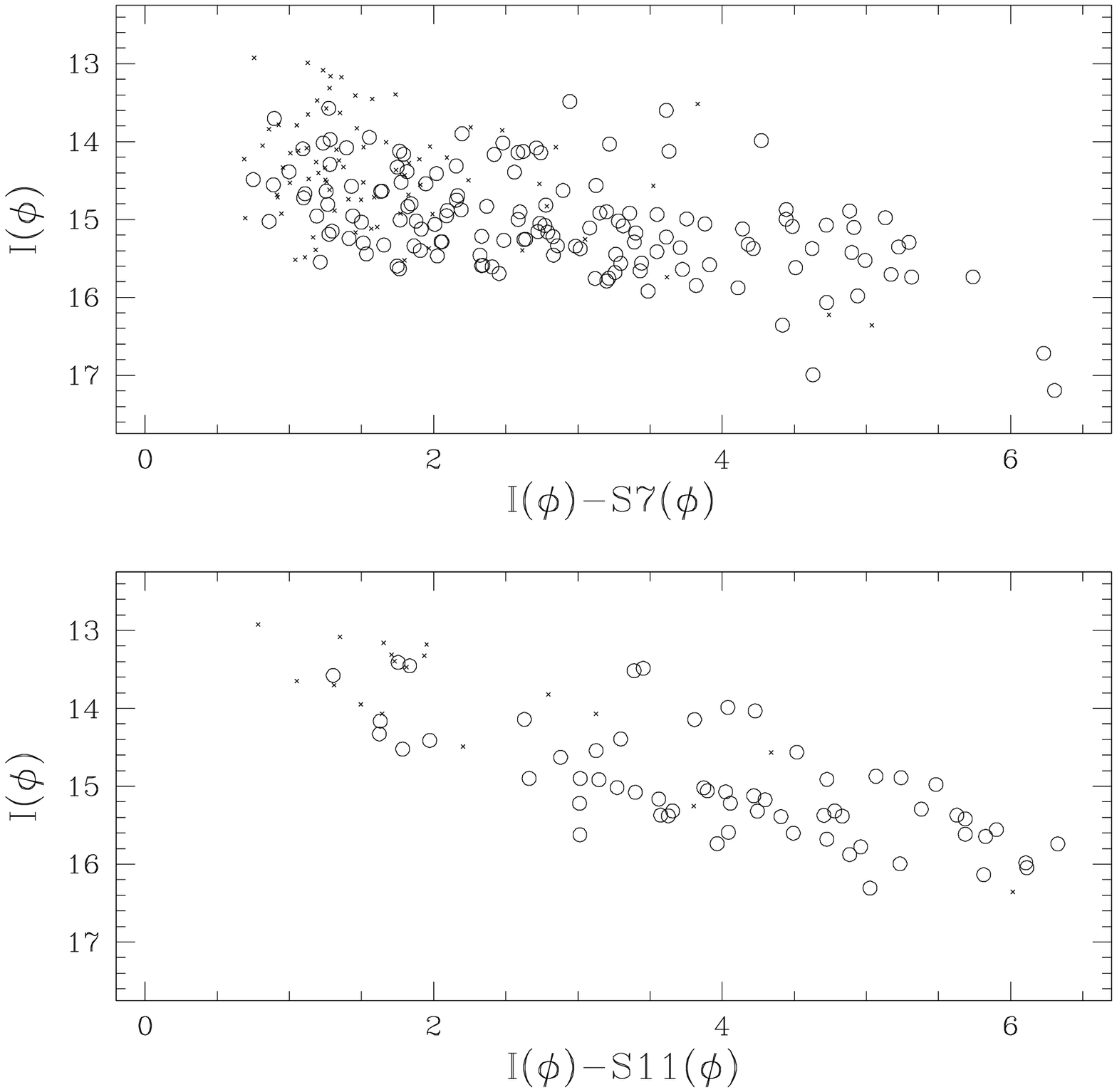}
  }
  \vspace{0cm}
  \caption{Left panel shows the random-phase CMD for the $I(\phi)-N3(\phi)$ colour. The dashed lines are the adopted colour cut at $0.5< I(\phi)-N3(\phi) < 1.7$. Right panels show the random-phase CMD for the $I(\phi)-S7(\phi)$ and $I(\phi)-S11(\phi)$ colours. Error bars are omitted for clarity. }
  \label{rawcmd}
\end{figure*}

It can be seen from upper left panel of Figure \ref{rawcolor} that colours with $I(\phi)-N3(\phi)<2.0$ show a periodic behavior. This is because most of these {\it AKARI} sources are indeed matched to the Cepheids, and hence this periodic behavior originates from the pulsational behavior of Cepheids. In contrast, the colours with $I(\phi)-S7(\phi)<2.0$ did not show a clear periodic behavior\footnote{Even if the $I(\phi)-S7(\phi)<2.0$ colours show a very weak periodic behavior, the number of possible truly matched {\it AKARI} sources is small for a meaningful analysis of the P-L relation in $S7$ band.}, and the periodic behavior is totally absent for $I(\phi)-S11(\phi)<2.0$ colours. The lower panels of Figure \ref{rawcolor} suggest that the matched {\it AKARI} sources with $I(\phi)-AKARI(\phi)>2.0$ in general have a larger separation: this further supports the conclusion that these {\it AKARI} sources are falsely matched to the OGLE-III Cepheids. Therefore, we only consider the matched {\it AKARI} sources in $N3$ band in this paper with the following selection criteria (based on Figure \ref{rawcolor}):

\begin{eqnarray}
\ \ \mathrm{Separation}  <  0.7 \mbox{arc-second} , \nonumber \\
0.5  <  I(\phi)-N3(\phi)  <  1.7. \nonumber 
\end{eqnarray} 

\ni These selection criteria are represented as a dashed box in the lower left panel of Figure \ref{rawcolor}. Figure \ref{rawcmd} shows the random-phase colour magnitude diagram (CMD) for the matched {\it AKARI} sources. Sources that satisfy the above selection criteria appears to define an instability strip occupied by Cepheids.

\section{The Random-Phase Correction}

The methodology for random-phase correction in the $JHK$ bands has been developed by \citet{nik04}, \citet{sos05} and \citet{maj08} in order to derive mean magnitudes from single epoch 2MASS data. All of these random-phase correction methods have their own advantages and disadvantages. The \citet{nik04} method is straight forward to apply, but their approach ignores the contribution of the amplitudes in individual Cepheids \citep{sos05}. The technique developed by \citet{sos05} and \citet{maj08} takes the amplitudes of the light curves into account. However, their method also relies on the existing light curve templates (for example in $JHK$ bands) and/or the amplitude ratios between the $JHK$ band and the $V$ (or $I$) band. 

In our case, the amplitude ratio between $N3$ and $V$ (or $I$) band is an unknown. Furthermore, there are no Cepheids with full light curves in the $N3$ band to develop the light curve template. We therefore need a modified approach to derive the random phase correction for our single-epoch $N3$ data. 

\subsection{The Methodology}

Following \citet{sos05}, we define a normalized light curve at phase $\phi$ as:

\begin{eqnarray}
T(\phi) & \equiv & \frac{m(\phi) - <m>}{A}, 
\end{eqnarray} 

\ni where $<m>$ and $A$ is the mean magnitude and amplitude of the light curve in a particular band, respectively, and $\phi$ is the phase calculated from equation (1). The function $T(\phi)$ can be modeled as an $n$-order Fourier series \citep{nik04,sos05}:

\begin{eqnarray}
T(\phi) & = & \sum_{j=1}^{n} [a_j\cos (2\pi j\phi) + b_j\sin (2\pi j\phi)].
\end{eqnarray}

\ni Re-arrange equation (2) such that $m=<m>+A\times T(\phi)$. If we denote the amplitude ratio between $N3$ band and $I$ (or $V$) band as $A_{N3}/A_I=C$, then we have $m=<m>+A_I C\times T(\phi)$. The amplitude ratio $C$ can be absorbed into the Fourier coefficients $a_j$ and $b_j$ in equation (3). Furthermore, the mean magnitude $<m>$ can be replaced by the definition of P-L relation, $<m>=\alpha\log(P)+\beta$. We then obtain a linear equation which can be solved with a standard least squares method:

\begin{eqnarray}
m & = & \alpha\log(P)+\beta \nonumber \\
  &   & +A_I \times \sum_{j=1}^{n} [a'_j\cos (2\pi j\phi) + b'_j\sin (2\pi j\phi)],
\end{eqnarray}

\ni where $a'_j=Ca_j$ and $b'_j=Cb_j$. Dropping the $A_I$ and $C$ term from equation (4), we recover the expression that is similar to the one given in \citet{nik04}:

\begin{eqnarray}
m = \alpha\log(P)+\beta+\sum_{j=1}^{n} [a_j\cos (2\pi j\phi) + b_j\sin (2\pi j\phi)].
\end{eqnarray}

\subsection{Test with the $K$ Band Data}

The performance of equation (4) and (5), with the random-phase correction term, to solve for the P-L relation can be tested with $K$ band data taken from \citet{per04}. We first matched the $88$ Cepheids \citep[after removing HV 883, HV 2447, HV 2883 and HV 12765, as in][]{per04} with OGLE-III Cepheid catalog from \citet{sos08} that has both the $I$ band amplitudes and $t_0$ information. There are $46$ Cepheids found to be matched between these two catalogs. Each of these Cepheids has $\sim19$ to $\sim30$ $K$ band data points per light curve.

\begin{figure}
  \centering 
  \epsfxsize=8.3cm{\epsfbox{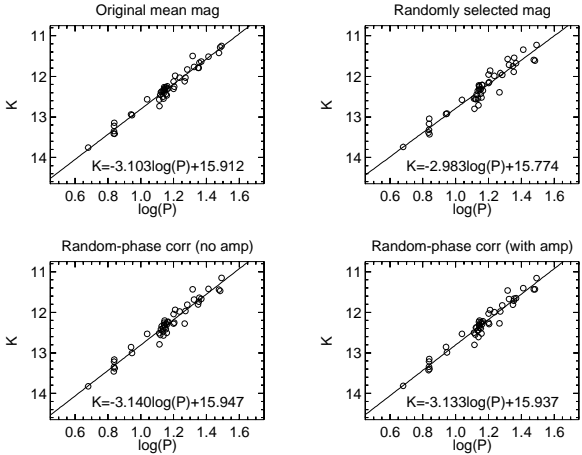}}
  \caption{An example of the comparison for the derived P-L relations in various cases. Upper left panel shows the P-L relation from using the mean magnitudes. Upper right panel shows the fitted P-L relation if the single epoch $K$ band data is randomly selected, without any random-phase correction. Lower left and right panels show the resulted P-L relations if equation (5) and (4) is employed, respectively. In both cases an $n=2$ order Fourier expansion is adopted. The randomly selected data presented here only represent one of the many runs we have tried.}
  \label{k_pltest}
\end{figure}

\begin{figure}
  \centering 
  \epsfxsize=8.0cm{\epsfbox{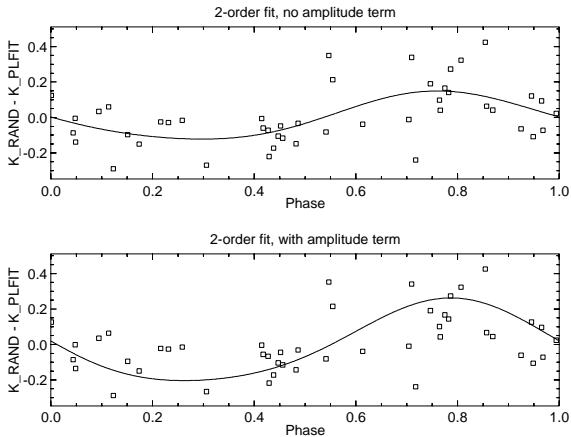}}
  \caption{Comparison of the fitted $T(\phi)$ function for the data presented in Figure \ref{k_pltest}, without (upper panel) and with (lower panel) the amplitude term. $K_{RAND}$ is the randomly selected single epoch data, and $K_{PLFIT}=\alpha\log(P)+\beta$, where $\alpha$ and $\beta$ are fitted results from equation (4) or (5).}
  \label{t_pltest}
\end{figure}

For each Cepheid, we randomly select one data point from the $K$ band light curves to represent the single epoch observation. We then fit the randomly selected data in three cases: (a) P-L relation without random-phase correction (i.e., $K_{\mathrm{random}}=\alpha\log[P]+\beta$ only); (b) P-L relation with random-phase correction from equation (4) that includes an amplitude term; and (c) P-L relation with random-phase correction from equation (5) without the amplitude term. Note that there is no extinction correction and/or photometric transformation applied to these $K$ band magnitudes, since the purpose here is {\it not} deriving the $K$ band P-L relations, but to test the random phase correction. We also fit the P-L relation using the mean magnitudes from \citet{per04} for these $46$ Cepheids in order to compare with the P-L relations derived from single epoch data. Figure \ref{k_pltest} presents the comparison of the P-L relations in these four cases from one run of the randomly selected single epoch $K$ band data. Figure \ref{t_pltest} shows the corresponding fitted $T(\phi)$ function without and with the amplitude term.   

We have repeated the above test procedure $5000$ times to built up the statistics for comparing the P-L from randomly selected data, either with or without the random-phase corrections, to the 'true' P-L relation derived from mean magnitudes. Histograms were constructed for these $5000$ runs when performing such comparisons, and a simple Gaussian function in the form of $A_Ge^{-(\Delta-m_G)^2/2s_G^2}$ was fitted to these histograms (where $\Delta$ is the difference of the slopes or zero-points of fitted P-L relations). If only using the randomly selected single epoch data, without any random-phase correction, then $m_G=0.017$ and $s_G=0.093$ for the comparison of the slopes and $m_G=-0.011$ and $s_G=0.113$ for the comparison of the zero-points of the fitted P-L relations. 

\begin{table}
  \centering
  \caption{Summary of the fitted Gaussian to the histograms for P-L slopes.}
  \label{tab_slope}
  \begin{tabular}{lcccc} \hline
     & \multicolumn{2}{c}{Without Amplitude} & \multicolumn{2}{c}{With Amplitude} \\
   $n$ & $m_G$ & $s_G$ & $m_G$ & $s_G$ \\
    \hline 
    1 &$-0.005$ & 0.057 &$-0.007$ & 0.049 \\
    2 &$-0.007$ & 0.061 &$-0.010$ & 0.052 \\
    3 &$-0.004$ & 0.064 &$-0.009$ & 0.058 \\
    4 &$-0.004$ & 0.071 &$-0.010$ & 0.059 \\
    5 &$-0.004$ & 0.076 &$-0.011$ & 0.067 \\
    6 &$-0.005$ & 0.079 &$-0.012$ & 0.071 \\ 
    \hline
  \end{tabular}
\end{table}

\begin{table}
  \centering
  \caption{Summary of the fitted Gaussian to the histograms for P-L zero-points.}
  \label{tab_zp}
  \begin{tabular}{lcccc} \hline
     & \multicolumn{2}{c}{Without Amplitude} & \multicolumn{2}{c}{With Amplitude} \\
   $n$ & $m_G$ & $s_G$ & $m_G$ & $s_G$ \\
    \hline 
    1 & $0.013$ & 0.062 & $0.015$ & 0.054 \\
    2 & $0.014$ & 0.066 & $0.018$ & 0.059 \\
    3 & $0.011$ & 0.074 & $0.017$ & 0.060 \\
    4 & $0.011$ & 0.080 & $0.019$ & 0.069 \\
    5 & $0.011$ & 0.079 & $0.021$ & 0.074 \\
    6 & $0.011$ & 0.091 & $0.023$ & 0.079 \\
    \hline
  \end{tabular}
\end{table}

\begin{figure*}
  \vspace{0cm}
  \hbox{\hspace{0.5cm}
    \epsfxsize=8.0cm \epsfbox{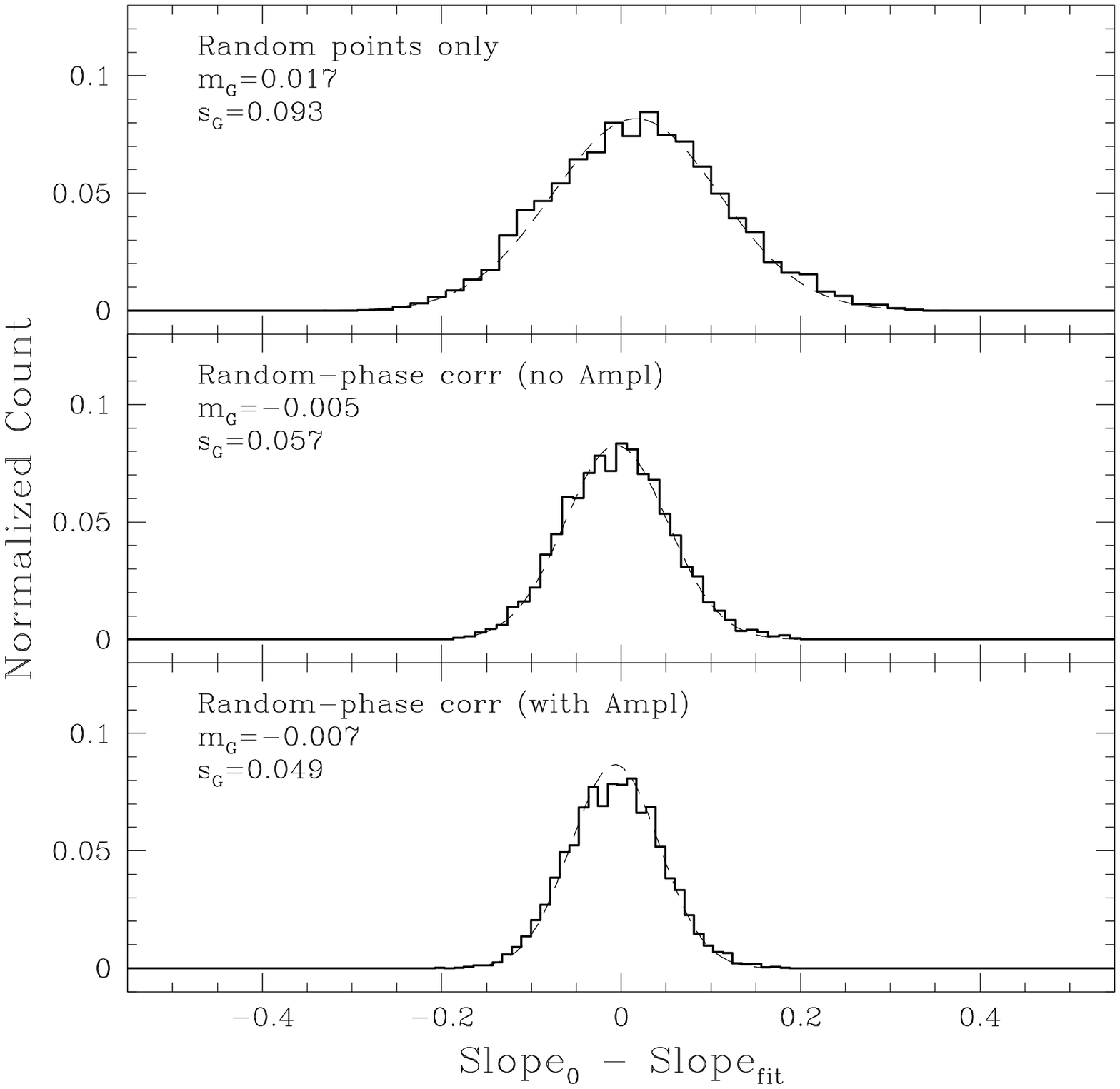}
    \epsfxsize=8.0cm \epsfbox{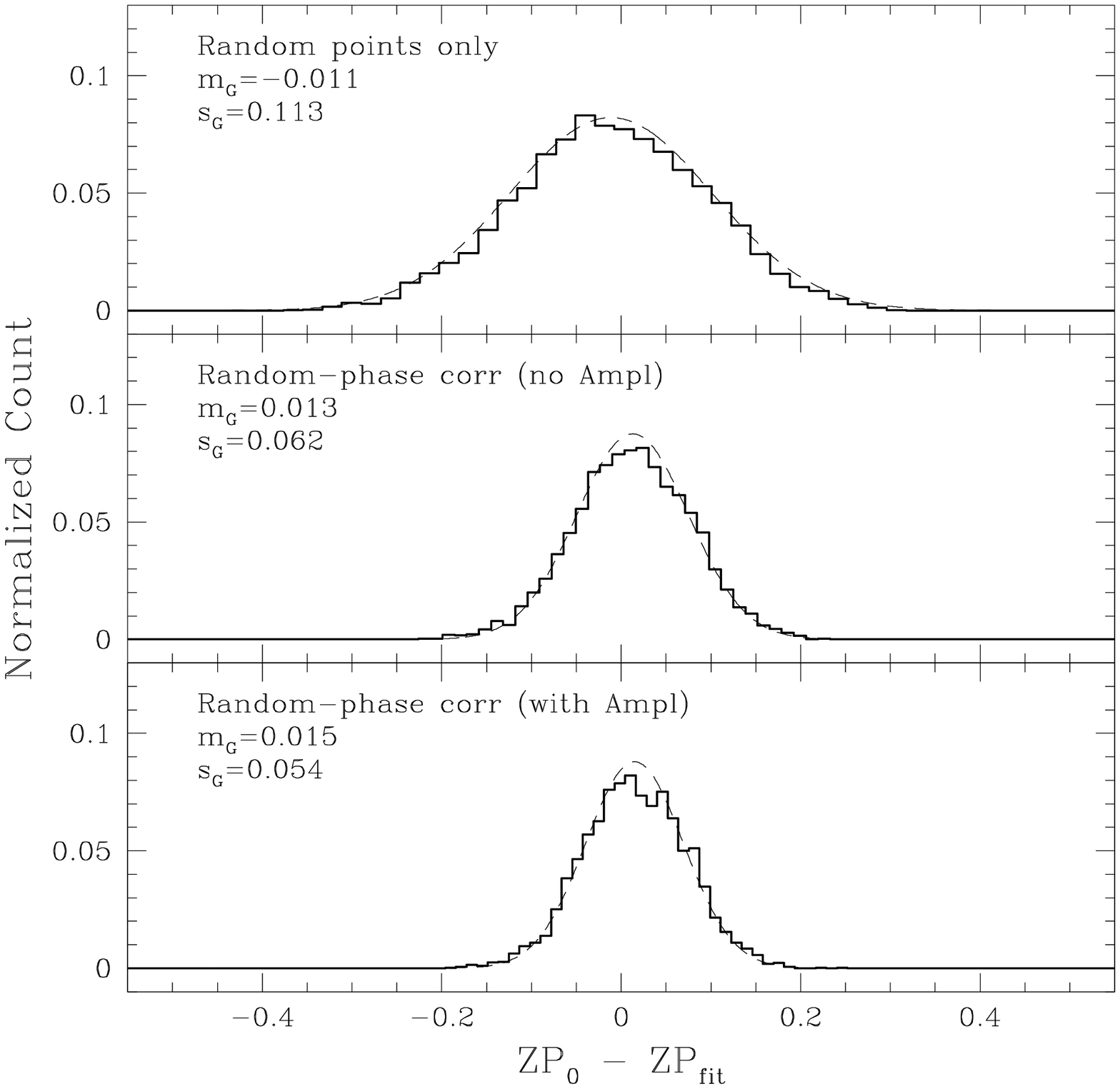}
  }
  \vspace{0cm}
  \caption{Histograms for the comparison of the P-L slopes (left panel) and P-L zero-points (right panel). In each cases, the fitted P-L relations from randomly selected single epoch data ($\mathrm{Slope}_{\mathrm{fit}}$ and $\mathrm{ZP}_{\mathrm{fit}}$) were compare to the P-L relations from using the mean magnitudes ($\mathrm{Slope}_0$ and $\mathrm{ZP}_0$). Upper panels are for the randomly selected data points only, without any random-phase correction. Middle and lower panels show the histograms if random-phase corrections with $n=1$ from equation (5) and (4) are applied to the data, respectively. The dashed curves are the fitted Gaussian function to the histograms. The fitted values of $m_G$ and $s_G$ from the Gaussian are given in the upper-left corner of each panels.}
  \label{histogram}
\end{figure*}

We also ran our test procedure and comparisons for $n=1$ to $n=6$ in the Fourier series when applying the random-phase corrections. Table \ref{tab_slope} and \ref{tab_zp} summarize the fitted values of $m_G$ and $s_G$ from the histograms with different $n$ for the P-L slopes and zero-points, respectively. Several results can be immediately seen from these Tables:

\begin{itemize}
\item Bias of the P-L slopes, $m_G$, has decreased from $0.017$ if random-phase correction is applied to the randomly selected single epoch data. Random-phase correction without the amplitude term, i.e. equation (5), seems to provide a better result.

\item There is not much improvement for the bias in the P-L zero-points when using the random-phase corrections. Furthermore, Random-phase corrections with the amplitude term (equation [4]) seems to increase this bias.

\item The dispersion of the histogram, $s_G$, has been improved by $\sim15$\% to $\sim50$\% when the random-phase correction is included in fitting the P-L relation. Random-phase correction with the amplitude term gives a smaller dispersion than the correction without the amplitude term.

\end{itemize}

\ni Overall, Table \ref{tab_slope} and \ref{tab_zp} suggest that random-phase correction from equation (4) with $n=1$ in the Fourier series gives the best fit results. Figure \ref{histogram} displays the histograms with $n=1$ case. The improvement of including the random-phase correction in fitting the P-L relations can be clearly seen in this Figure. We therefore include the random-phase correction with $n=1$ when deriving the $N3$ band P-L relation in the next section.

We also compared the estimated mean magnitudes from the random-phase correction ($m-T[\phi]$) to the 'true' mean magnitudes from \citet{per04}. A Gaussian fit to the histograms for the difference in mean magnitudes reveal that there is no bias introduced from the random-phase correction. The error of the estimated mean magnitudes is about $0.06$mag, for both of the random-phase correction that either do or do not include the amplitude term.

\section{The $N3$ Band P-L Relation}

There are $338$ matched {\it AKARI} sources in the $N3$ band left after the selection criteria described in Section 2.1\footnote{The $N3$ photometry of these Cepheids will be available at the SIMBAD CDS database ({\tt http://simbad.u-strasbg.fr/simbad/}) once the revised catalog (Kato et al. 2010) is published.}. We further remove one Cepheid with $\log(P)=0.066$ to avoid contamination from overtone Cepheids. To remove any possible outliers presented in the data, we employed an iterative $2.5\sigma$-clipping procedure to fit equation (4) and (5), that include the random-phase corrections, when deriving the P-L relation. The resulted P-L relations (for $327$ Cepheids) are:

\begin{description}
\item No random-phase correction: \newline
  $N3  =  -3.198(\pm0.046)\log(P)+15.816(\pm0.031)$, $\sigma=0.161$
\item From equation (5):  \newline
  $N3  =  -3.189(\pm0.042)\log(P)+15.811(\pm0.029)$, $\sigma=0.148$
\item From equation (4): \newline
  $N3  =  -3.193(\pm0.043)\log(P)+15.813(\pm0.029)$, $\sigma=0.149$
\end{description}

\begin{figure}
  \centering 
  \epsfxsize=8.3cm{\epsfbox{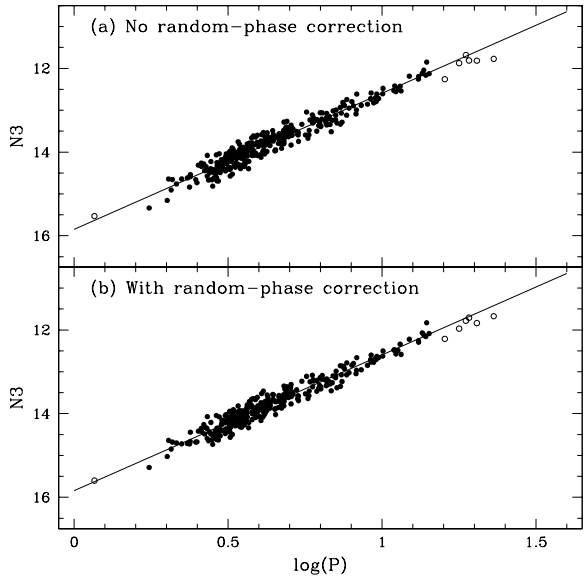}}
  \caption{The final P-L relations in the $N3$ band, for using the single epoch random phase data only (upper panel) and with the inclusion of random-phase correction (lower panel). The P-L relations derived from equation from (4) and (5) are identical, hence only one of them is plotted here. Open circles are the excluded Cepheids (see text for details). Error bars are omitted for clarity.}
  \label{pl_final}
\end{figure}

\begin{figure}
  \centering 
  \epsfxsize=8.3cm{\epsfbox{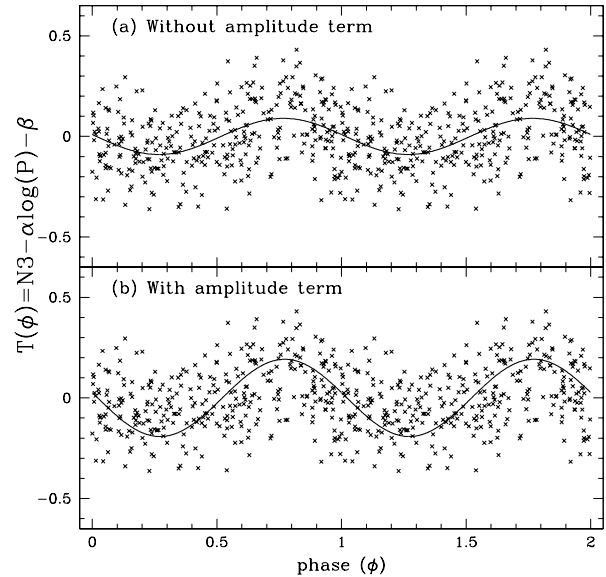}}
  \caption{The corresponding $T(\phi)$ function for the Cepheids presented in Figure \ref{pl_final}. Upper and lower panel shows the $T(\phi)$ function without and with the amplitude term, respectively.}
  \label{t_final}
\end{figure}

\ni The derived P-L relation without the random-phase correction agrees well to the P-L relations that include the random-phase corrections. This further supports the suggestion that random-phase corrections may not be needed for deriving the mid-infrared P-L relations, due to the expected small amplitudes in those bands \citep{nge08}. However, the dispersion of the P-L relation ($\sigma$) decreases by $\sim7$\% if the random-phase correction is included. The P-L relations derived from either using equation (4) or (5) are almost identical, suggesting the amplitude term may not need to be included in the random-phase correction.  

Since the $N3$ band is closely matched to the {\it Spitzer's} $3.6\mu\mathrm{m}$ band, the slope of the $N3$ P-L relation is expected to close to the $3.6\mu\mathrm{m}$ P-L relations. The slopes from the latest determined $3.6\mu\mathrm{m}$ P-L relations are: $-3.253(\pm0.010)$ from \citet{nge09} and $-3.40(\pm0.02)$ from \citet{mad09}. The $N3$ P-L relation seems to be shallower than the $3.6\mu\mathrm{m}$ P-L relations. Plotting out the $N3$ P-L relation reveals that the $6$ Cepheids with $\log(P)>1.2$ appear to be (systematically) fainter than what is expected from a ridge line of the P-L relation (see also the upper panel in Figure \ref{pl_raw}). This can cause the P-L relation to appear shallower. Note that the saturation limit at $N3$ band is $7.8$mag \citep{ita08}, which cannot account for the appearing zero-point offset of the photometry for these $6$ long period Cepheids. If we excluded these long period Cepheids, then the fitted P-L relations (for $321$ Cepheids) are: 

\begin{figure*}
  \vspace{0cm}
  \hbox{\hspace{0.5cm}
    \epsfxsize=8.0cm \epsfbox{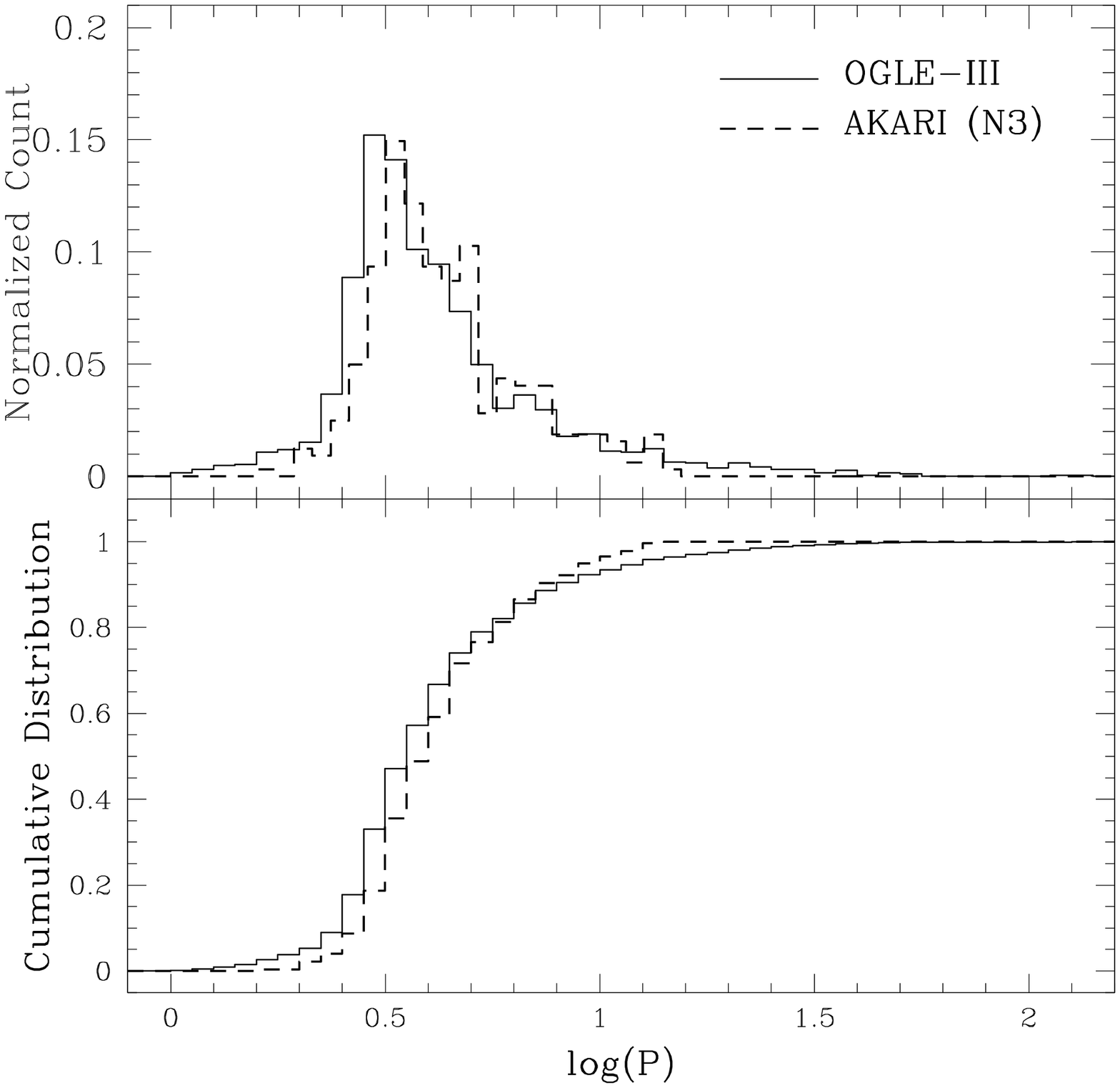}
    \epsfxsize=8.0cm \epsfbox{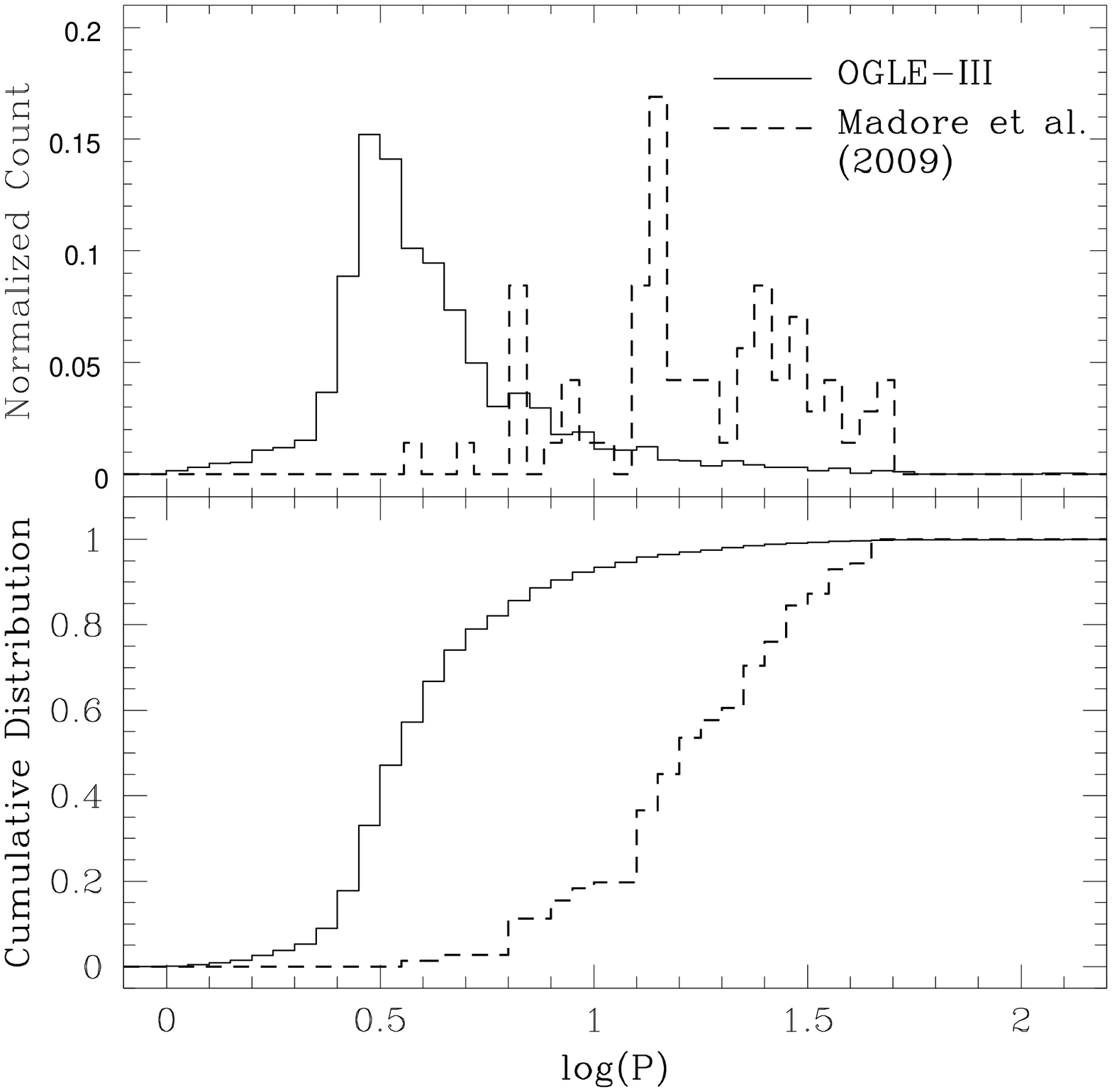}
  }
  \vspace{0cm}
  \caption{Comparison of the period distribution for the {\it AKARI's} $N3$ sample we used to derive the P-L relation (upper left panel) and the Cepheid sample in \citet[][upper right panel]{mad09} to the full sample of OGLE-III Cepheids. Lower panels show the corresponding cumulative period distribution of these samples.}
  \label{periodcompare}
\end{figure*}

\begin{description}
\item No random-phase correction: \newline
  $N3  =  -3.254(\pm0.051)\log(P)+15.850(\pm0.034)$, $\sigma=0.160$
\item From equation (5):  \newline
  $N3  =  -3.241(\pm0.047)\log(P)+15.842(\pm0.031)$, $\sigma=0.147$
\item From equation (4): \newline
  $N3  =  -3.246(\pm0.047)\log(P)+15.844(\pm0.031)$, $\sigma=0.149$
\end{description}

\ni Figure \ref{pl_final} presents the final $N3$ P-L relations, and Figure \ref{t_final} shows the corresponding $T(\phi)$ as a function of phase for the final adopted Cepheids. Note that the slope of the $N3$ P-L relation without the random-phase correction is identical to the {\it Spitzer's} $3.6\mu\mathrm{m}$ band P-L relation from \citet{nge09}, which the random-phase corrections were not applied to the SAGE data. We took the P-L relation derived from equation (4) as our final adopted $N3$ P-L relation:

\begin{equation}
N3 =  -3.246\log(P) + 15.844
\end{equation} 

\subsection{Comparison to the $3.6\mu\mathrm{m}$ P-L Relation}

The slope of $-3.246$ from the $N3$ P-L relation is closer to the $3.6\mu\mathrm{m}$ P-L slope given in \citet[][$-3.253$]{nge09} rather than the slope from \citet[][$-3.40$]{mad09}. A possible explanation for this discrepancy is the different sample size used in \citet{nge08} and \citet{mad09}. The sample in \citet{mad09} consist of $71$ Cepheids with most of them have period longer than $10$ days. The mean period of their sample is $<\log(P)>=1.242$, in contrast to OGLE-III ($0.622$) and the $N3$ sample used in this paper ($0.641$). In the upper left panel of Figure \ref{periodcompare}, we compare the period distribution of the final {\it AKARI's} $N3$ sample that we used to derive the P-L relation to the $\sim1800$ Cepheid sample from OGLE-III catalog. It can be seen that both samples show a consistent period distribution. This is further supported from the cumulative distribution of the two samples, as presented in lower left panel of Figure \ref{periodcompare}. In contrast, the period distribution from \citet{mad09} is inconsistent with the OGLE-III sample, as shown in the right panel of Figure \ref{periodcompare}. 

\begin{table}
  \centering
  \caption{The effect of period cuts on the $N3$ P-L slope.}
  \label{tab_pcut}
  \begin{tabular}{lccc} \hline
   $\log(P_{\mbox{cut}})$ & Slope & $N$ & $<\log(P)>$  \\
    \hline 
    0.50  & $-3.203\pm0.056$ & 261 & 0.688 \\
    0.55  & $-3.156\pm0.065$ & 207 & 0.731 \\
    0.60  & $-3.205\pm0.074$ & 163 & 0.774 \\
    0.65  & $-3.231\pm0.085$ & 131 & 0.810 \\
    0.70  & $-3.293\pm0.113$ & 91  & 0.869 \\
    0.75  & $-3.399\pm0.135$ & 75  & 0.902 \\
    0.80  & $-3.524\pm0.151$ & 60  & 0.934 \\
    0.85  & $-3.468\pm0.209$ & 43  & 0.978 \\
    \hline
  \end{tabular}
\end{table}

\citet{nei09b} found that the slope of the $3.6\mu\mathrm{m}$ P-L relation becomes steeper, and comparable to the slope given in \citet{mad09}, when a period cut of $\log(P_{\mbox{cut}})=1.05$ is applied to the SAGE data from \citet{nge09}. In Table \ref{tab_pcut}, we present the slopes for the $N3$ P-L relation when various period cut ($\log[P_{\mbox{cut}}]$) is applied to the $N3$ data, together with the number of Cepheids left in the sample ($N$) and the mean period of the sample ($<\log[P]>$). From this Table, the slope of the $N3$ P-L relation and the number of Cepheid in the sample are almost same as the results given in \citet{mad09}, if a period cut of $\log(P_{\mbox{cut}})\sim0.75$ is applied. 

\begin{figure}
  \centering 
  \epsfxsize=8.0cm{\epsfbox{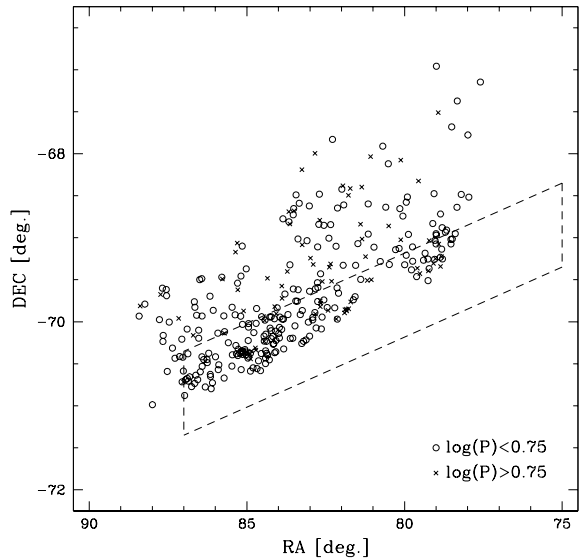}}
  \caption{Location of the Cepheids in our final adopted sample, separated by $\log(P)=0.75$. The dashed box roughly outline the location of the LMC bar region.}
  \label{location}
\end{figure}

Since some of the {\it AKARI} sources located at the LMC bar region (where the stellar density is higher), could the problem of crowding affects the slope of the $N3$ P-L relations as seen in Table \ref{tab_pcut}? The locations of the Cepheids from our final adopted sample are shown in Figure \ref{location}, where the dashed box represents the LMC bar region. We divided our sample into two sub-samples: one within the bar region and one outside the bar region. In Table \ref{tab_region}, the P-L slopes derived from these two sub-samples were compared, together with their mean period and the number of Cepheids separated at $\log(P)=0.75$. Due to influence of the crowding on the photometry, especially for the short period Cepheids, the P-L slope for the sub-sample from the bar region is expected to be shallower. However, this is not seen from the results given in Table \ref{tab_region}, which suggest that crowding within the bar region does not affect the {\it AKARI's} photometry. Furthermore, the slopes from the two regions are consistent with each other. We have shifted, expanded or contracted the bar region as shown in Figure \ref{location}, the results still hold with similar slopes presented in Table \ref{tab_region}.

The structure of the mid-infrared P-L relation has been found to be modified by the presence of Circumstellar Envelopes \citep{nei09,nei09b}, like those discovered by \citet{ker06}. It is argued that the envelopes are formed by a Cepheid wind where dust forms at a large distance from the star and causes an IR excess \citep{nei08}. This IR excess changes the structure of the mid-infrared P-L relation, by increasing the observed mid-infrared brightness of Cepheids. The IR excess is found to be important in two regimes, the first is for Cepheids with $\log(P)<0.7$, and for long-period Cepheids. The IR excess of short-period Cepheids is significant and makes the slope of the P-L relation appear more shallow than it actually is. For long-period Cepheids, the IR excess causes the slope to become steeper. The observed $3.6\mu\mathrm{m}$ P-L relation for $\log(P) > 1.0$ is $-3.23$ \citep{nge09}, and when the IR excess is removed the slope is predicted to be $-3.19$. For increased period cuts, the predicted P-L relations tend to have a shallower slope. It was also noted that the observed P-L relation tends to steepen with increased period cut and shown that the samples in \citet{nge09} and \citet{mad09} are consistent. This suggests that a fraction of the Cepheids in \citet{mad09} sample could be affected by IR excess due to mass loss.

\citet{fre08b} suggested that the slopes of the mid-infrared P-L relation should reach an asymptotic value that can be predicted from the slope of the period-radius (P-R) relation. If ignored the temperature term at the mid-infrared, the predicted luminosity is $L\propto R^2$, and hence $m\equiv -2.5\log(L)\propto -5\log(R)$. Adopting the observed P-R slope of $0.68$ from \citet{gie99}, then the predicted P-L slopes in the mid-infrared is $-3.40$. This value is closer to the slopes found in \citet{fre08b} and \citet{mad09}\footnote{If we accept this argument, our slopes of $\sim -3.25$ from the $N3$ and $3.6\mu\mathrm{m}$ P-L relations are consistent with the theoretical P-R slope from \citet[][$0.653$ from the non-canonical models for LMC metallicity at $Z=0.008$]{bon98}.}. However, \citet{nei09b} argue that the temperature term cannot be ignored, since the mid-infrared luminosity is at the Rayleigh-Jeans tail of the black-body function ($L\propto R^2T_{\mbox{eff}}$). If including the temperature term, then the predicted P-L slopes at the mid-infrared is $\sim-3.23$ \citep[for more details, see][]{nei09b}, which is closer to the slopes found in the $N3$ band from this paper and in \citet{nge09}.

\section{Conclusion}

In this paper, we matched the {\it AKARI's IRC} catalog with OGLE-III LMC Cepheid catalog. This allows us to derive the P-L relation in mid-infrared bands. We have applied a conservative cut on the matched sources based on the colour information (at random phase) and the separation between the matched sources in the two catalogs. Only the matched sources in $N3$ band have sufficient number of Cepheids for deriving the P-L relation after applying these selection criteria. 

\begin{table}
  \centering
  \caption{Comparison of $N3$ P-L slopes from two regions.}
  \label{tab_region}
  \begin{tabular}{lcc} \hline
    & In the LMC bar & Outside the LMC bar  \\
    \hline 
    P-L slope         &  $-3.275\pm0.079$ & $-3.240\pm0.056$ \\
    $<\log(P)>$      & 0.627  & 0.658 \\
    $N_{\log(P)<0.75}$&  135 &  111 \\
    $N_{\log(P)>0.75}$&  31  &   44 \\
    \hline
  \end{tabular}
\end{table}

We have employed a modified random-phase correction based on the methodology adopted from \citet{nik04} and \citet{sos05}. The correction function was modeled with a lower order Fourier series which either include or without the amplitude term. The $K$ band light curve data from \citet{per04} was used to test the fitting procedure including the random-phase correction. It was found that random-phase correction with $n=1$ in the Fourier series provides the best fit results. 

The derived $N3$ band P-L relation with random-phase correction is consistent with the P-L relation without the random-phase correction. However, the dispersion of the P-L relation was improved by $\sim7$\% when the random-phase correction is applied to the single epoch $N3$ band data. The slope of the $N3$ P-L relation, which is independent of the SAGE data, is in good agreement with the $3.6\mu\mathrm{m}$ P-L relation found in \citet{nge09}. 

\section*{Acknowledgments}

This research is based on observations with {\it AKARI}, a JAXA project with the participation of ESA. We thank the {\it AKARI IRC} LMC survey team for early access to the data, and Nancy Evans for constructive discussion. We would also like to thank the referee, Barry F. Madore, for useful comments. This work is partly supported by a Grant-in-Aid for Scientific Research (A) No.~18204014 from Japan Society for the Promotion of Science. CCN thank the funding from National Science Council (of Taiwan) under the contract NSC 98-2112-M-008-013-MY3.

\label{lastpage}

\end{document}